# Discrete metasurface for extreme sound transmission through water-air interface

Shao-Cong Zhang[1], Hong-Tao Zhou[1], Xiao-Tong Gong[1], Yan-Feng Wang[1,]*  and Yue-Sheng Wang[1,2]

[1]Department of Mechanics, School of Mechanical Engineering, Tianjin University, Tianjin 300350, China

[2]Department of Mechanics, Beijing Jiaotong University, Beijing 100044, China

*Corresponding author, Email: wangyanfeng@tju.edu.cn

**Abstract:** The mismatch of acoustic impedance at water-air interface can lead a low transmitted sound energy. In this paper, we propose a discrete metasurface for extreme sound transmission based on the impedance matching theory. By employing topology optimization, discrete unit cells with different aspect ratios are designed with unitary sound transmission. The unit cell of continuous metasurface is also obtained for comparison. After analyzing the wide-angle performance of discrete unit cells, samples of both discrete and continuous metasurfaces are fabricated. Sound transmission enhancement of discrete metasurface is clearly measured compared to the bare water-air interface. And the amplitude is relatively larger than that of the continuous sample. Experimental results are in general agreement with numerical ones when viscosity of the sample is considered. Furthermore, the frequency shifts between experiment and simulation are attributed to the random immersion of unit cells for discrete metasurface and the bending of continuous metasurface, respectively. The present work suggests an alternative way for improving the efficiency of water-air acoustic communication.

## 1. Introduction

Hydro-acoustic communication is currently the main means of communication between land and ocean[1]. However, the ratio of the acoustic impedance of water and air is up to 3600, leading to almost 30 dB transmission loss of sound through the interface[2-4]. The impedance mismatch at water-air interface thus highly limits the communication efficiency of radar, sonar, and other acoustic applications[5-7]. To handle this problem, a specific media placed at the water-air interface is required. Both impedance and thickness of this media should satisfy the impedance matching conditions[2]. However, though many efforts have been made, it is difficult to find an appropriate media, whose impedance is 60 times than that of air[8]. Recently, the concept of acoustic metasurface has provided an alternative way for solving the problem.

Acoustic metasurface is a kind of artificial structures consisting of subwavelength unit-cells[9]. It can exhibit abnormal functions of acoustic waves, such as abnormal refraction[10], mode conversion [11-13], focusing[14-16], carpet cloaking[17], etc. Specially, metasurfaces are used to manipulate acoustic waves across the water-air interface. The investigations are classified into two parts:

First, a number of studies are aiming to achieve extreme sound transmission across the water-air interface. The structure may consist of loaded membrane[18], Helmholtz resonators[19] above the water-air interface, or bubbles[20-22] immersed in the water. Bok et al[18] realize unitary sound transmission at water-air interface using the resonance of the tensioned membrane units with central mass. Zhang et al[19] design a composite waveguide for extreme sound transmission through the water-air interface and the impedance is adjusted by changing the geometrical parameters. Lee et al[21] discussed the effects of the radius of the underwater bubble and the distance of the bubble to the water-air interface on the impedance matching. Moreover, Huang et al[22] experimentally realized high sound transmission through the interface based on the resonance of the underwater bubbles. Nevertheless, the design of these metasurfaces is usually limited at low frequency.

Second, there are some recent investigations focusing on the wavefront manipulation of acoustic waves across the water-air interface. Liu et al[23] constructed a hydroacoustic metasurface with acoustic-focusing effect and enhanced sound transmission. Combining the single-phase unitary-transmission metasurface and phase-modulated acoustic metasurface, Zhou et al[24] realized focusing different vortex fields across water-air interface.

It is noted that all these reported metasurfaces are continuous. They should be fully replaced even when part of them fail in use. In this paper, we propose a discrete metasurface for enhanced sound transmission across water-air interface. Unit-cells of the discrete metasurface are designed by digging holes in an isotropic elastic solid using topology optimization. The design scheme has no restriction on the frequency range. Numerical calculations are implemented by Finite Element Method. Unit-cells with different aspect ratio are obtained. For comparison, the unit-cell of continuous metasurface is also designed using the same method. Extreme sound transmissions are achieved based on the impedance matching theory. The transmission performances of the unit-cells for oblique incidence and the metasurfaces assembled by these units are investigated. Discrete and continuous metasurfaces are then fabricated by 3D printing. Enhanced sound transmissions are clearly measured across the water-air interface. Finally, the frequency shifts between experiment and simulation are analyzed by considering the random immersion of the discrete metasurface and the bending of the continuous metasurface.

## 2. The impedance matching theory

To realize extreme transmission across the water-air interface, we first consider an isotropic transformer with length $w$ and thickness $l$ placed at interface[25], as illustrated in Figure 1(a). When harmonic plane wave with angular frequency $\omega$ is incident from water to air, the acoustic fields and the velocity can be expressed as

$$P_i = P_{ia}e^{j(wt-k_w y)}, P_r = P_{ra}e^{j(wt+k_w y)}, P_{Mt} = P_{Mta}e^{j(wt-k_M y)}, P_r = P_{Mra}e^{j(wt+k_M y)}, P_t = P_{ta}e^{j(wt-k_a y)} \quad (1)$$

$$v_i = \frac{P_i}{\rho_w c_w}, v_r = \frac{P_r}{\rho_w c_w}, v_{Mt} = \frac{P_{Mt}}{\rho_M c_M}, v_{Mr} = \frac{P_{Mr}}{\rho_M c_M}, v_t = \frac{P_t}{\rho_a c_a} \quad (2)$$

where $P_i$ and $P_r$ is the incident wave and reflected wave in the water below y=-$l$, $v_i$ and $v_r$ is the corresponding velocity, respectively; $P_t$ is the transmitted wave in the air above y=0, $v_t$ is the corresponding velocity; $P_{Mt}$ and $P_{Mr}$ are the transmitted wave and the reflected wave in the transformer between y=-$l$ and y=0, respectively. $P_{ia}, P_{ra}, P_{Mta}, P_{Mra}$ and $P_{ta}$ represent the amplitude of $P_i, P_r, P_{Mt}, P_{Mr}$ and $P_t$, respectively; j= $\sqrt{-1}$; $k_w$ ($\rho_w$, $c_w$), $k_a$ ($\rho_a$, $c_a$) and $k_M$ ($\rho_M$, $c_M$) represent the wavenumber (density, sound velocity) of water, air and the transformer, respectively.

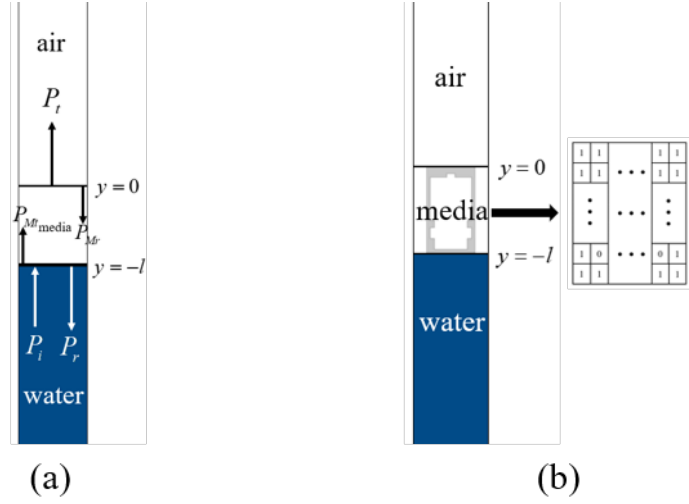

Fig. 1. Schematics of sound transmission across the water-air interface with (a) the effective isotropic transformer, or (b) the optimized unit cell made of a single-phase solid.

On the interface between two different media, the sound pressure and velocity of the transmitted wave should be continuous, i.e.,

$$P_{ia}e^{k_w l} + P_{ra}e^{-k_w l} = P_{Mta}e^{k_M l} + P_{Mra}e^{-k_M l}, \quad y=-l \tag{3}$$

$$v_i e^{-k_w l} - v_r e^{k_w l} = v_{Mt} e^{-k_M l} - v_{Mr} e^{k_M l}, \quad y=-l \tag{4}$$

$$P_{Mta} + P_{Mra} = P_{ta}, \quad y=0 \tag{5}$$

$$v_{Mt} - v_{Mr} = v_t, \quad y=0 \tag{6}$$

Combining Eqs. (3)-(6), the sound transmission T from water to air can be given by

$$T = \frac{|P_t|^2/2\rho_a c_a}{|P_i|^2/2\rho_w c_w} = \frac{4Z_a Z_w}{(Z_a+Z_w)^2 \cos^2 k_M l + (Z_M + \frac{Z_a Z_w}{Z_M})^2 \sin^2 k_M l} \tag{7}$$

where $Z_w = \rho_w c_w, Z_a = \rho_a c_a$ and $Z_M = \rho_M c_M$ represent the impedance of water, air and the media, respectively. Extreme sound transmission T=1 can be obtained when $\cos k_M l$ =0 and $Z_M^2 = Z_a Z_w$. Then we have:

$$l = \frac{(2n-1)}{4}\lambda \tag{8}$$

$$Z_M = \sqrt{Z_a Z_w} \tag{9}$$

where $\lambda = 2\pi/k_M$ is the wavelength in the transformer.

## 3. Topology optimization method

In this paper, topology optimization and genetic algorithm are employed to explore the transformer with extreme sound transmission across the water-air interface. The single-phase transformer is made of epoxy with density $\rho$=1180kg·m$^{-3}$, the Young's modulus E=2.65GPa and the Poisson's ratio $\upsilon$ = 0.41. The density and sound velocity of water and air is 1000 kg·m$^{-3}$ and 1500m/s, 1.21kg·m$^{-3}$ and 343 m/s, respectively. The operating frequency is chosen as $f$=10kHz.

The genetic algorithm is used to maximize the sound transmission of the unit cell. And the optimization formulation can be expressed as

$$\text{Find:} \quad \rho(i,j) = 0 \text{ or } 1 \ (i=1,2,\dots,K; j=1,2,\dots,L) \tag{10}$$

$$\text{Minimum:} \quad y(f) = 1 - T \tag{11}$$

$$\text{Subject to:} \quad \rho(i,j) = 1 \ (i=1 \text{ or } K \text{ and } j=1 \text{ or } L) \tag{12}$$

$$\text{Max}(K \text{ or } L) = 24 \tag{13}$$

where $y(f)$ is the fitness function for the target; K and L are the number of pixels in the design domain shown in Figure 1(b), K varies from 8 to 24 and L=24. Each pixel is given a density $\rho_i$ of '0' or '1', representing for air (or solid). The outermost side of the design domain is set as '1' to improve the speed of optimization. Meanwhile, considering the symmetry of the water and air regions, as well as the form of plane wave propagation, the unit cell is designed to be y-axis symmetrical. Besides, the "filter"[26] is used to remove isolated elements and fill isolated voids.

The evolutionary iteration procedure of the topological optimization is describing as follows: first, the initial population ($K_i$) with size (S) is randomly generated, at the same time the evolutionary generation (G) is set as 0; Second, calculate the fitness function ($y_i$) of the current population by using COMSOL Multiphysics; Third, if $y_i$ is smaller than the minimum relative error (Er=$10^{-4}$), break the loop and print $K_i$, otherwise use "tournament selection" to select new population ($K_{i+1}$) of population size (S). Fourth, perform crossover and mutation on the new population and repeat the 2nd~4th step.

## 4. Results and Discussion

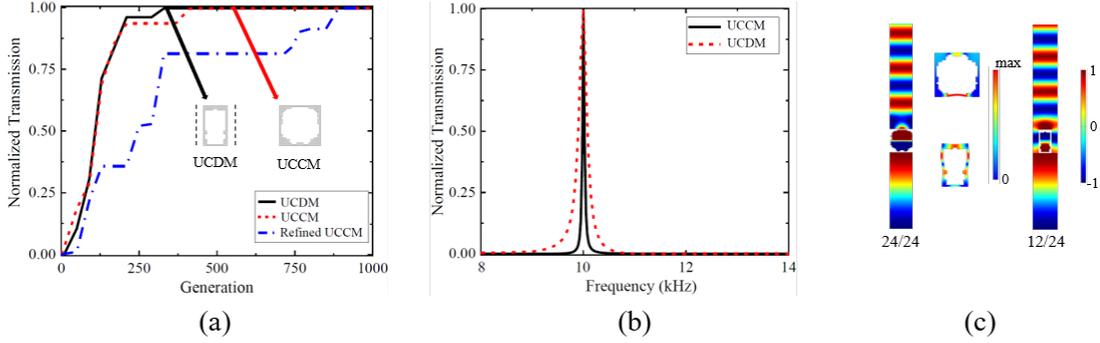

Fig. 2. The sound transmitted performance of the two kinds of unit cells. (a) The evolution of the unit cells with the increase of the generation. (b) The transmission of the unit cells with frequency. (c) The distributions of displacement and acoustic pressure of the unit cells under extreme transmission.

### 1) Unit cells with extreme sound transmission

By using topology optimization described in section 3, unit cell discrete metasurface (UCDM) with extreme transmission and aspect ratio of 1/2 of is obtained, as shown in Figure 2(a). The aspect ratio is defined as K/L. The unit cell of continuous metasurface (UCCM) is also obtained when K/L=1. It clarifies the differences in the number of evolutions of different cells. It shows that the topology optimization of these two kinds of cells starts from a random structure in G=1, which cannot satisfy the fitness function in Eq.(11) and ends before G=450. It is indicated that the type of the cell can hardly influence the generations of evolution. Figure 2(b) shows the transmitted sound pressure of these two unit cells at operating frequency. It can be seen that both of them have narrow band of frequency for high transmission. The deformation and transmitted sound field of these two cells at resonant frequency are illustrated in Fig. 2 (c), respectively. It can be clearly seen that the UCCM have two kinds of flexural vibrations at the water-air interface and in the air, which is the first order flexural vibration and the third order flexural vibration along y axis, respectively. However, that of the UCDM is different. It contains not only the first order flexural vibration and the third order flexural vibration along y axis, but also the second order flexural vibration along x axis in the air. This is because the UCCM has different periodic condition from the UCDM. At the same time, there exists two kinds of phase in the transmitted sound pressure.

Then extreme transmissions are also obtained for unit cells with different aspect ratio, as shown in Figure 3(a). It can be observed that the sound transmissions of these unit cells are close to 1. Figure 3(b) is the deformation and acoustic pressure of each cell. It depicts that there are two kinds of phases for the unit cells. And the phase difference is π.

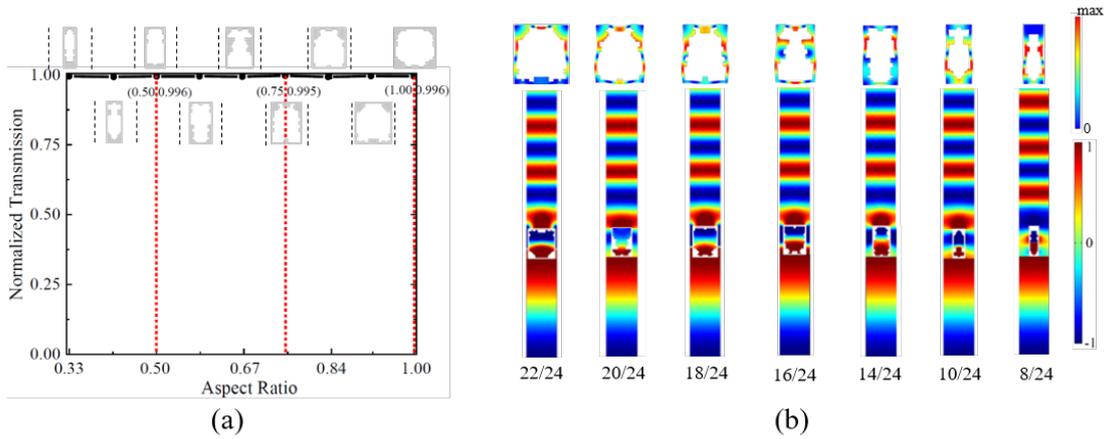

Fig. 3. The sound transmitted performance of the unit cells with different aspect ratio. (a) The transmission of the unit cells with different aspect ratio. (b) The corresponding distributions of displacement and acoustic pressure for the unit cells with extreme transmission.

The equivalent parameters of the unit cells with different aspect ratio can be obtained using transfer matrix method [27]. The results are shown in Table 1 with equivalent speed $c'$, equivalent density $\rho'$, equivalent impedance $Z'$ and equivalent wavelength $\lambda'$ of UCDM, respectively. It can be obtained that all of the unit cells have similar normalized equivalent impedance, which satisfies the impedance matching condition in Eq. (9). At the same time, the difference of the equivalent wavelength of the unit cell also suggests a phase change of π according to Eq. (8), consistent with the results in Figs. 2(c) and 3(b).

Table 1 Equivalent parameters of the discrete unit cells with different aspect ratios

| Aspect ratio | $c'$ (m/s) | $\rho'$ (kg/m$^{-3}$) | $Z'/Z_M$ | $l/\lambda'$ |
|---|---|---|---|---|
| 8/24 | 309 | 79.74 | 0.99 | 0.74 |
| 10/24 | 918 | 26.92 | 0.99 | 0.25 |
| 12/24 | 930 | 26.65 | 0.99 | 0.25 |
| 14/24 | 921 | 26.87 | 0.99 | 0.25 |
| 16/24 | 927 | 26.52 | 0.99 | 0.25 |
| 18/24 | 935 | 26.41 | 0.99 | 0.25 |
| 20/24 | 915 | 26.94 | 0.99 | 0.25 |
| 22/24 | 923 | 26.71 | 0.99 | 0.25 |
| 24/24 | 305 | 81.02 | 0.99 | 0.74 |

**2) Wide-angle performance of the discrete unit cell**

In this part, the transmission performance of UCDM under incident wave with different angles is discussed. As shown in Figure 3(a), the sound wave is incident from water side with incident angle $\theta_i$ and transmitted to air side with refracted angle $\theta_t$, According to Snell's Law, we can get

$$k_w \sin\theta_i = k_a \sin\theta_t \tag{14}$$

In this occasion, the sound transmission coefficient in the normal direction of the unit cell is

$$T' = \frac{k_a^2 - k_w^2 \sin^2\theta_i}{k_a^2 \cos^2\theta_i} T \tag{15}$$

By using Eq. (15), the effects of incident angle on the sound transmission of the UCDM and UCCM are calculated and shown in Figs. 4(a) and (c), respectively. It can be clearly seen that the sound transmission of UCDM is close to 1 for most incident angles. However, high transmission of UCCM can be obtained only when the incident angle is around 0°. The refracted acoustic fields of UCDM and UCCM are plotted in Figs. 4(b) and (d). The refracted angle of both UCDM and UCCM varies within a small range when incident angle is changed from 0° to 90°. This could be explained by Eq. (14). Since $k_w$ is about five times larger than $k_a$, $\theta_t$ should be no larger than 18° even for 90° incidence. Meanwhile, both unit cells can realize high transmission when the incident angle is less than 18°. However, the color of the pressure distribution for UCCM fades away when the incident angle is over 36°. While the UCDM has relatively high transmission when the incident angle is no more than 72°. So the UCDM has a better wide-angle performance than the UCCM.

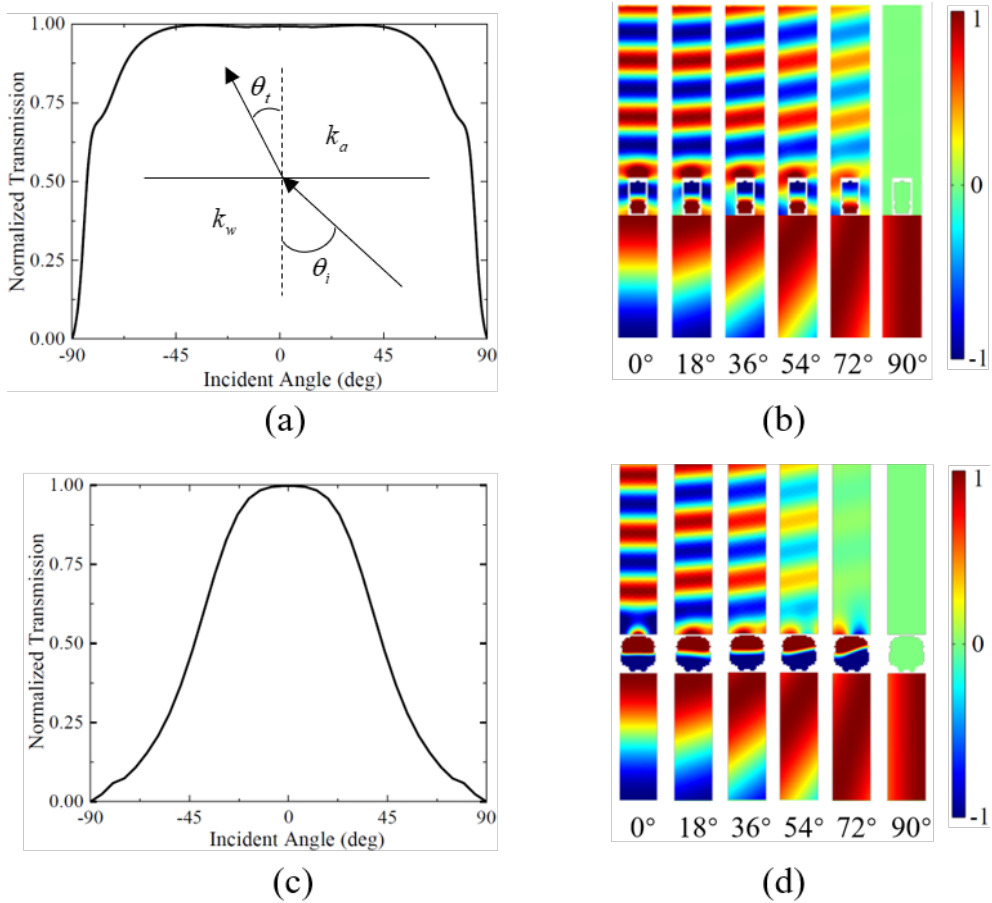

Fig. 4. The transmitted performances of the UCDM and UCCM for different oblique incidences. Panels (a) and (c) show the sound transmission of UCDM or UCCM with different incident angles. The inset in panel (a) is the generalized Snell's law. Panels (b) and (d) illustrates the sound pressure fields of UCDM or UCCM with incident angle varying from 0° to 90°.

## 3) Discrete metasurfaces with extreme transmission

In this part, we construct discrete and continuous metasurfaces using the unit cells mentioned above. Each metasurface consists of 20 unit cells. Gaussian-type plane wave is used as the excitation in the water. The transmitted sound fields of the discrete and continuous metasurfaces at the operating frequency are shown in Figs.5 (a) and (b). Plane waves are observed in air for both metasurfaces. However, the energy at the transmitted side of the discontinuous metasurface is more concentrated than that of the continuous one. This is due to the wide-angle performance of UCDM. As shown in Fig. 3 (c), the transmission of the UCDM is smaller in the normal direction. So a large amount of the energy is refracted to other direction, leading to the low concertation of the transmitted energy of the continuous metasurface.

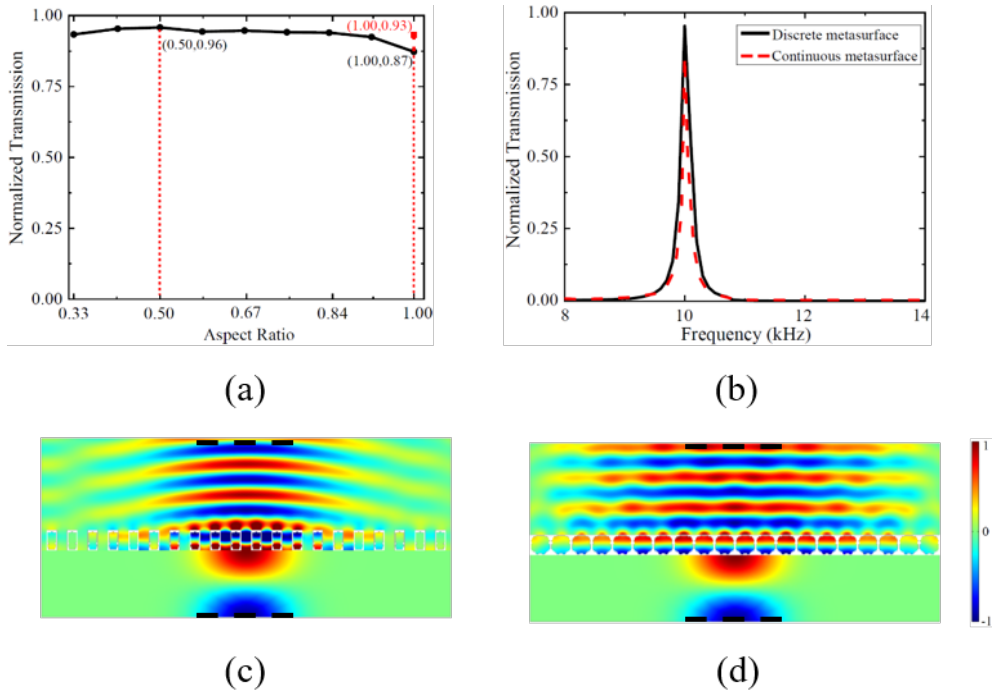

Fig. 5. The sound transmitted performance of the discrete and continuous metasurfaces across the water-air interface. Panel (a) shows the sound transmission of the discrete metasurface with different aspect ratio at the operating frequency. Panel (b) presents variation of sound transmission for the discrete and continuous metasurfaces with frequency. Panels (c) and (d) illustrate the distribution of sound pressure of the discrete and continuous metasurfaces, respectively.

To quantitively characterize the transmission properties, the transmitted pressure is collected on a line with length equal to wave source, as marked in Figs. 5(c) and (d). After integration of sound pressure in these two lines, the sound transmission of the metasurface can be calculated by using Eq. (7). The corresponding results are illustrated in Fig. 5(c). It is found that the result is similar to that of the unit-cells in Fig. 2 (b). Both metasurfaces have a narrow band of frequency for high transmission. We further evaluate the transmission properties of discrete metasurface with other aspect ratio, as shown in Fig. 5(d). It is found that all the discrete metasurface have relative larger transmission compared to the continuous one.

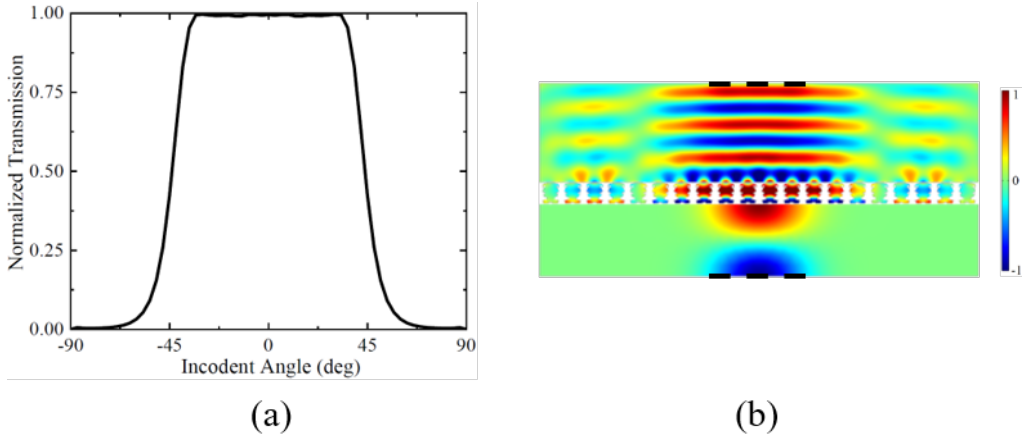

Fig. 6. The wide-angle performance of the refined UCCM. (a) The transmission of the refined UCCM for various incidences. (b) The transmitted sound field of the continuous metasurface with refined UCCM.

It should be noted that the continuous metasurface can have high transmission by improving the wide-angle performance. This could be realized by limiting UCCM with high transmission for a relatively large range of incident angle, e.g., 0°, 20° and 35°. And the corresponding fitness function can be modified as

$$y^{'}(f) = 1 - \alpha T(0°) - \beta T(20°) - \gamma T(35°) \qquad (16)$$

where $\alpha=0.5$, $\beta=0.4$ and $\gamma=0.1$ are the weighting coefficients. Fig. 6(a) shows that the sound transmission through the refined unit cell for incident angle between -90° to 90°. It is found that unitary transmission is obtained for the incident angle from -33° to 33°.

The corresponding range of incident angle is larger than that in Fig. 4(c). However, the convergence of the unit cell is much slower than UCDM due to the strong constraints of the target, as illustrated in Fig. 2(a). The transmitted sound field of the UCCM is shown in Fig. 6(b). It is found that the energy at the transmitted side of the new continuous metasurface is more concentrated compared with Fig. 4(d). The sound transmission of the new continuous metasurface is 0.93, much close to that of the discrete metasurface, see the red point marked in Fig. 4(a). This indicates that the wide-angle performance of unit cell gives rise to the high transmission of the metasurface.

**4) Experimental verification**

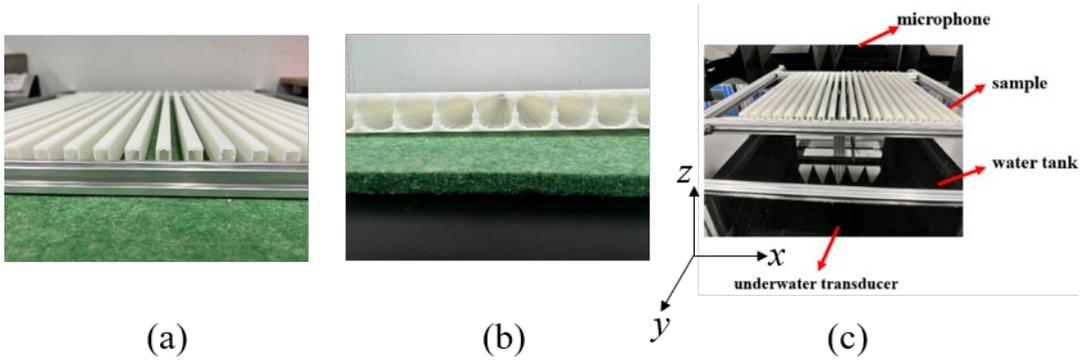

Fig. 7. Photograph for the samples of metasurfaces and experimental setups. (a) The discrete metasurface made of 20 units with aspect ratio of 1/2. (b) The continuous metasurface made of 20 units. (c) The experimental setups.

In order to verify the high transmission of discrete and continuous metasurfaces, we fabricate the samples with 20 unit cells by 3D printing, as shown in Figs. 7(a) and (b). They are stretched by 50 cm in the non-periodic direction. Fig. 7(c) shows the experiment setups for measuring transmitted acoustic field of metasurfaces. Harmonic wave source with frequency from 8kHz to 14kHz is generated using B&K PLUS Labshop software. It is output by the Input and Output Module (B&K Type 3160-A-042). After being amplified by power amplifier (Krohn-Hite 7500 and B&K Type 2573), the source is played by the underwater transducer with a diameter of 10cm placed on the bottom of the water tank (dimensions 1.6m×1.5m×0.8m). The discrete and continuous metasurfaces are placed on a square steel frame with a width of 65cm and a length of 50cm, respectively. The frame with metasurface is put on the water-air interface. And the transmitted signal is received by the microphone (B&K Type 4939). The scanning area is 20cm×20cm above the center of the sample.

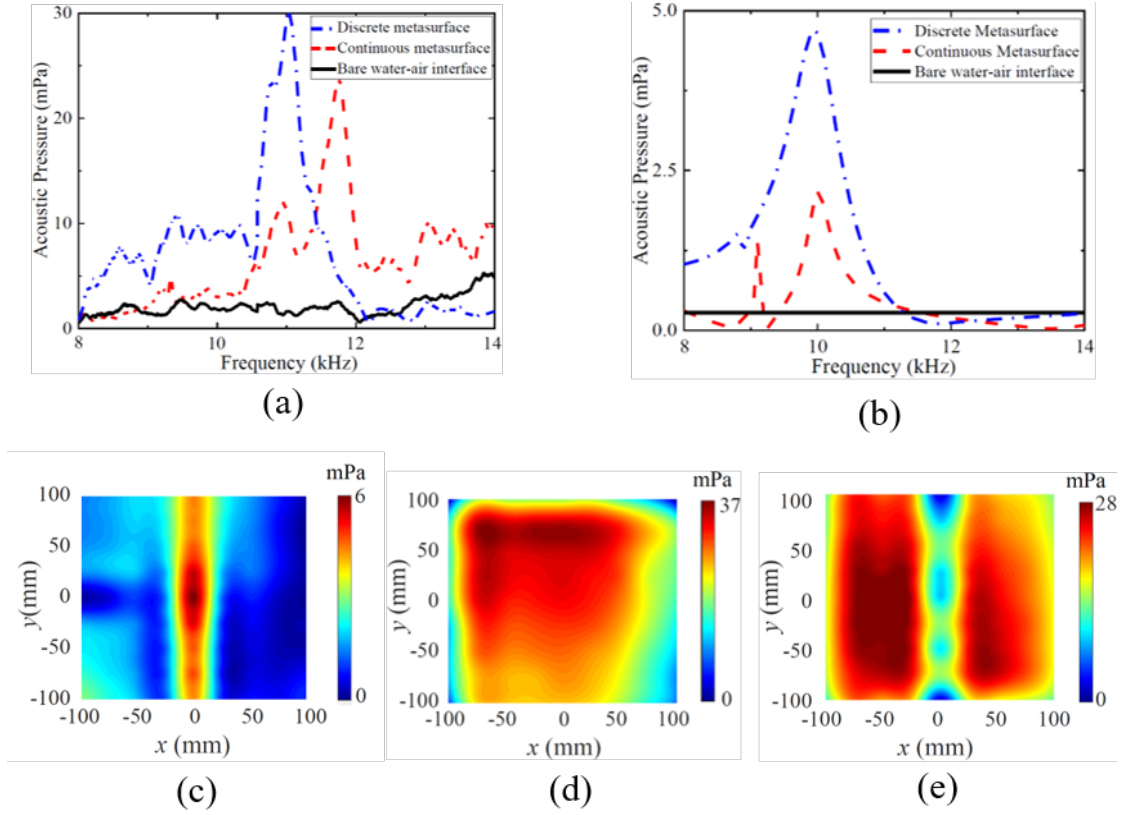

Fig. 8. Experimental results of the discrete and continuous metasurfaces. Panel (a) shows the measured sound pressure of the bare water-air interface, discrete and continuous metasurfaces. Panel (b) illustrates the simulated transmission with viscosity considered for the discrete and continuous metasurfaces. For comparison, the result for bare water-air interface is also presented. Panels (c)-(e) present the distributions of measured sound pressure in $x$-$y$ plane for the bare water-air interface at $f$=10kHz, discrete metasurface at $f$=10.9kHz and continuous metasurface at $f$=11.7kHz, respectively.

Figure 8(a) shows the average measured transmission of discrete and continuous metasurfaces. As a comparison, the transmission of bare water-air interface is also measured. It is observed that both discrete and continuous metasurfaces have relatively large transmission. The transmission of the discrete metasurface is almost 20 times larger than that of bare interface at 11.2kHz. And the result for continuous metasurface is 10 times larger compared to the bare interface at 11.8kHz. Figure 8(c)-(e) are the corresponding transmitted sound amplitude filed in *x-y* plane. It can be clearly seen that the amplitude of the sound pressure for the bare water-air interface much smaller than that for the discrete and continuous metasurfaces. And the amplitude distribution of sound for the discrete metasurface is more concentrated than that for the continuous one. These results are consistent with simulation in Figure 4(c) and (d).

Compared to Figure 4(b), the curve in Figure 8(a) is blunter. This might be resulted from the viscosity of the solid and fluid. We then use DMA to measure the viscosity of the sample, which is around 0.05E. The simulated result with viscosity included is shown in Figure 8(b). It can be observed that the sound transmission for the bare water-air interface is a constant in the considered frequency range. And it is much more smaller compared to those of the metasurfaces at $f$=10kHz. Meanwhile, the sound transmission of the discrete metasurface is about twice larger than that of the continuous metasurface at $f$=10kHz. These results are in good agreement with the measured results shown in Figure 8(a).

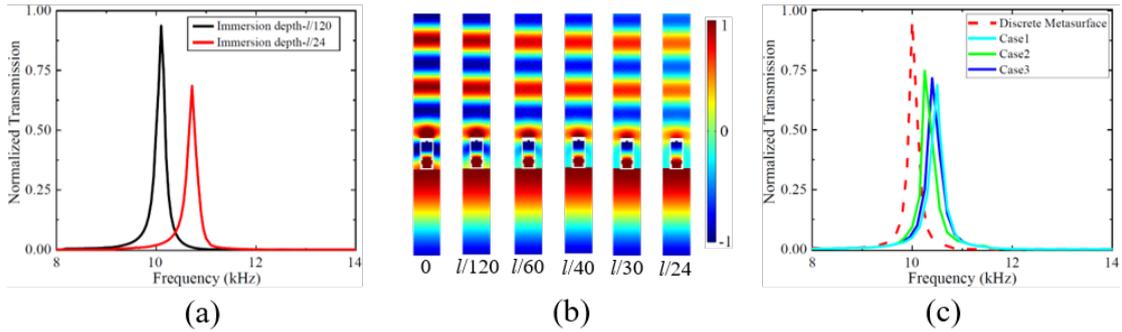

Fig. 9. The effect of random immersion depth of UCDMs on transmitted performance. Panel (a) shows the sound transmission of the UCDM with the immersion depth of $l/120$ and $l/24$. Panel (b) illustrates the transmitted sound filed of the UCDM at $f$=10kHz for different immersion depths. Panel (c) presents the sound transmissions of the discrete metasurfaces with/without random immersions.

However, clear frequency shift can be obtained between simulation and experiment for both metasurfaces. We attribute this to the random immersion of UCDM and the bending of the continuous metasurface, as we state below.

### 5) Effect of random immersion on transmission for discrete metasurface

In this part, we discussed the effect of random immersion of UCDMs on the transmission performance of the discrete metasurface. Fig. 9(a) shows the transmission of UCDM with immersion depths of $l/120$ and $l/24$, respectively. Right shift of the transmission peak is observed even for a small immersion. And the shift becomes further clear for a large immersion. The transmission amplitude is also decreased. This can also be observed from the pressure distribution at the transmission peak are shown in Fig. 9(b). Then we construct three discrete metasurfaces by UCDMs with random immersions, as illustrated in Fig. 9(c). The transmission peaks are shift to a larger frequency, which is consistent with the experiment in Fig. 9(a).

### 6) Effect of bending on the transmission for continuous metasurface

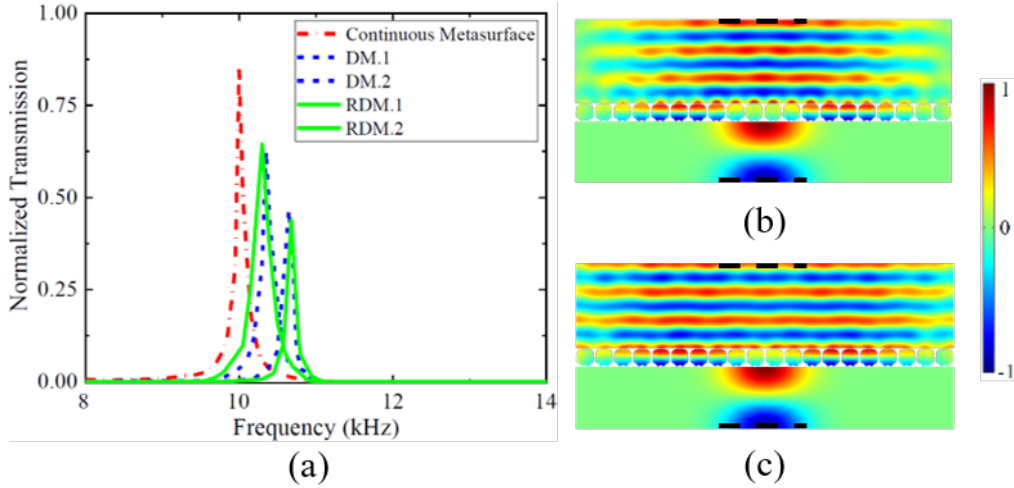

Fig. 10. The effect of bending on sound transmission and frequency shift for continuous metasurface. (a)The sound transmission of the continuous metasurface, DMs and RDMs. Panels (b) and (c) are the transmitted sound fields of the DMs with the maximum deformation of $l/120$ at $f=10.2$kHz and $l/24$ at $f=10.8$kHz, respectively.

The sample of continuous metasurface is bent after fabrication, even when it is restricted by the frame. A two-step simulation is conducted to analyze the effect of bending on the transmission performance of metasurface. The first is to calculate the static deformation of the metasurface. Static load along the normal direction is applied on the middle of the metasurface with two ends fixed. Considering the bending direction, two sets of deformed metasurface (DM, in the direction of wave propagation) and reverse deformed metasurface (RDM, in the opposite direction of wave propagation) are constructed, respectively. By changing the magnitude of the load, the maximum deformation of DM (or RDM) is $l/120$ and $l/24$, respectively. Then we evaluate the sound transmission performances of these metasurfaces. The corresponding result is shown in Figure 10(a). It is found that the transmission peaks of the deformed metasurfaces shift right and the amplitude decreases. The bigger the load is, the larger the deformation and the frequency shift, and the lower the transmission is. Figure 10(b)-(c) shows the transmitted sound field of the DM with different deformation at the corresponding peak frequency. Though the transmitted sound energy in Fig. 10(b) is more concentrated than that in Fig. 5(d), the transmission amplitude is smaller. It can also be observed form Figs. 10(b) and (c) that the transmitted pressure distribution for DM with larger deformation is more distributed, giving rise to a smaller transmission.

## 5. Conclusions

In this work, we propose discrete metasurfaces for enhanced sound transmission through the water-air interface by using topology optimization. For comparison, a continuous acoustic metasurface is also designed by the same optimization. Samples of discrete and continuous metasurfaces are then fabricated and measured. Pressure distributions of the samples are obtained from both simulations and experiment. The results show that enhanced sound transmission is obtained when the impedance matching conditions of the water, air and metasurface are satisfied. Although enhanced transmission is observed for both discrete and continuous metasurfaces in the simulations and experiment, the measured enhancement for the discrete metasurface is more pronounced due to the wide-angle performance of the discrete unit cell. This is also explained by the simulation when viscosity of the sample is considered. Meanwhile, the measured transmission peaks are shifted to higher frequencies compared to the simulated results. This is explained by the random immersion of the unit cells for discrete metasurface and the bending for the continuous metasurface.

In the future, the possibility to design uneven metasurfaces suitable for the uneven water-air metasurface will be explored. The present work will also be integrated with phase-modulated acoustic metasurface to realize abnormal wave functions through the water-air interface.

## Acknowledgements

This work is supported by National Natural Science Foundation of China (Grant Nos. 12072223, 12122207, 12021002 and 11991032). The first author also thanks Mr. Jiaxuan Weng for the valuable discussions and constructive suggestions.